\documentclass[aps,pra,twocolumn,superscriptaddress,shownopacs]{revtex4}

\usepackage{graphicx}
\usepackage{hyperref}
\usepackage{amsmath}
\usepackage{epstopdf}

\expandafter\let\csname equation*\endcsname\relax
\expandafter\let\csname endequation*\endcsname\relax 
\usepackage{amsmath,amssymb, color}
\usepackage{graphicx}
\newcommand{\bra}[1]{   \langle #1 |  }

\newcommand{\ket}[1]{  { | #1 \rangle}  }

\newcommand{\abs}[1] {  | #1 |  }

\newcommand{\ba}{\begin{eqnarray}}
\newcommand{\ea}{\end{eqnarray}}
\newcommand{\be}{\begin{equation}}
\newcommand{\ee}{\end{equation}}

\begin{document}

\title[Witnessing single-photon entanglement with local homodyne measurements]{Witnessing single-photon entanglement with local homodyne measurements: analytical bounds and robustness to losses}

\author{Melvyn Ho\footnotemark[2]\footnotetext{$^{\dagger}$ These authors contributed equally to the work.}}
\affiliation{Centre for Quantum Technologies, National University of Singapore, 3 Science Drive 2, Singapore 117543}
\affiliation{Department of Physics, University of Basel, CH-4056 Basel, Switzerland}
\author{Olivier Morin\footnotemark[2]}
\affiliation{Laboratoire Kastler Brossel, Universit\'{e}
Pierre et Marie Curie, Ecole Normale Sup\'{e}rieure, CNRS, 4 place
Jussieu, 75252 Paris Cedex 05, France}

\author{Jean-Daniel Bancal}
\affiliation{Centre for Quantum Technologies, National University of Singapore, 3 Science Drive 2, Singapore 117543}

\author{Nicolas Gisin}
\affiliation{Group of Applied Physics, University of Geneva, CH-1211 Geneva 4, Switzerland}

\author{Nicolas Sangouard}
\affiliation{Department of Physics, University of Basel, CH-4056 Basel, Switzerland}
\affiliation{Group of Applied Physics, University of Geneva, CH-1211 Geneva 4, Switzerland}

\author{Julien Laurat}
\email{julien.laurat@upmc.fr}

\affiliation{Laboratoire Kastler Brossel, Universit\'{e}
Pierre et Marie Curie, Ecole Normale Sup\'{e}rieure, CNRS, 4 place
Jussieu, 75252 Paris Cedex 05, France}

\begin{abstract}
Single-photon entanglement is one of the primary resources for quantum networks, including quantum repeater architectures. Such entanglement can be revealed with only local homodyne measurements through the entanglement witness presented in [Morin \textit{et al.} Phys. Rev. Lett. \textbf{110}, 130401 (2013)]. Here, we provide an extended analysis of this witness by introducing analytical bounds and by reporting measurements confirming its great robustness with regard to losses. This study highlights the potential of optical hybrid methods, where discrete entanglement is characterized through continuous-variable measurements.
\end{abstract}

\maketitle

\section{Introduction}
The optical hybrid approach of quantum information, which consists of mixing in a protocol both discrete and continuous degrees of freedom, has recently seen important developments. This includes advancements in quantum state engineering, state characterization and long-distance communication architectures \cite{Loock2011,Furusawa2011,Sangouard10,Brask10,hybrid1,hybrid2}. Based upon this approach, a witness for single-photon entanglement \cite{DLCZ, vanEnk05,Bjork01,HessmoPRL2004}, namely states of the form $|1\rangle_A|0\rangle_B+|0\rangle_A|1\rangle_B$ where $A$ and $B$ are two spatial modes sharing a delocalized single-photon, has been recently proposed and experimentally tested \cite{Morin2013}. It relies only on homodyne detections, i.e. on continuous quadrature measurements and not on photon counting, and offers significant advantages relative to other witnessing methods \cite{Babichev2004, Chou05, Laurat2007,LauratNJP}. Indeed, it does not require post-selection and does not assume knowledge of the underlying Hilbert space dimension (unlike most steering experiments~\cite{Bennet12}). Also, in contrast with other entanglement witnesses~\cite{Cavalcanti09}, it identifies the entanglement present specifically in the single-photon subspace. Finally, the measurements are operated only locally on the entangled modes, an important feature if applied to large scale networks \cite{Kimble08, Sangouard11}.

The witness presented in Ref. \cite{Morin2013} was built up on numerical arguments. In the present work, we extend its analysis by means of analytical calculations. The aim is to gain insight into the properties of the witness with respect to various practical imperfections. In particular, we investigate both theoretically and experimentally its robustness with regard to the channel loss or, equivalently, to imperfect single-photon states used as the initial resource for the entanglement generation. 
We demonstrate that even for a large admixture of vacuum, our witness reveals the presence of entanglement, confirming its suitability for use in realistic networks and entanglement distribution protocols where losses are inherent.

The paper is organized as follows. Section \ref{principle} first gives an overview of the single-photon entanglement witness based on local homodyne measurements. Then, in the case where the state only contains vacuum and single-photon components, i.e. the state lies within a qubit subspace,  the witness parameter is evaluated and compared to the separable bound. Symmetric and asymmetric channels are considered. In section \ref{bounding}, multiphoton components, which are critical in experimental realizations, are taken into account. We show how the witness is extended to this realistic case by experimentally bounding the Hilbert space and we then derive the effect of losses in the communication channels. This study leads to several expressions for the separable bound. The setup is presented in section \ref{manip} together with the experimental results. Conclusion is given in section \ref{Discussion}.

\section{Principle of the witness}\label{principle}
This section presents the principle of the single-photon entanglement witness, as proposed and demonstrated in \cite{Morin2013}, which relies only on local homodyne measurements. We then introduce the specific focus of this paper, i.e. the behavior of this witness in the presence of loss, coming equivalently from single-photon generated with non-unity efficiency or subsequent losses in the communication channels. In this section, the state is assumed to belong to the qubit subspace $\{|0\rangle,|1\rangle\}^{\otimes 2}$. This simplistic restriction allows us to understand the main features of the witness before generalizing the discussion to include multiphoton components.

\subsection{A Bell test scenario with local homodyning}
\begin{figure}[h!]
\centerline{
\includegraphics[width=0.95\linewidth,trim = 4.1cm 7.5cm 4.5cm 2cm, clip]{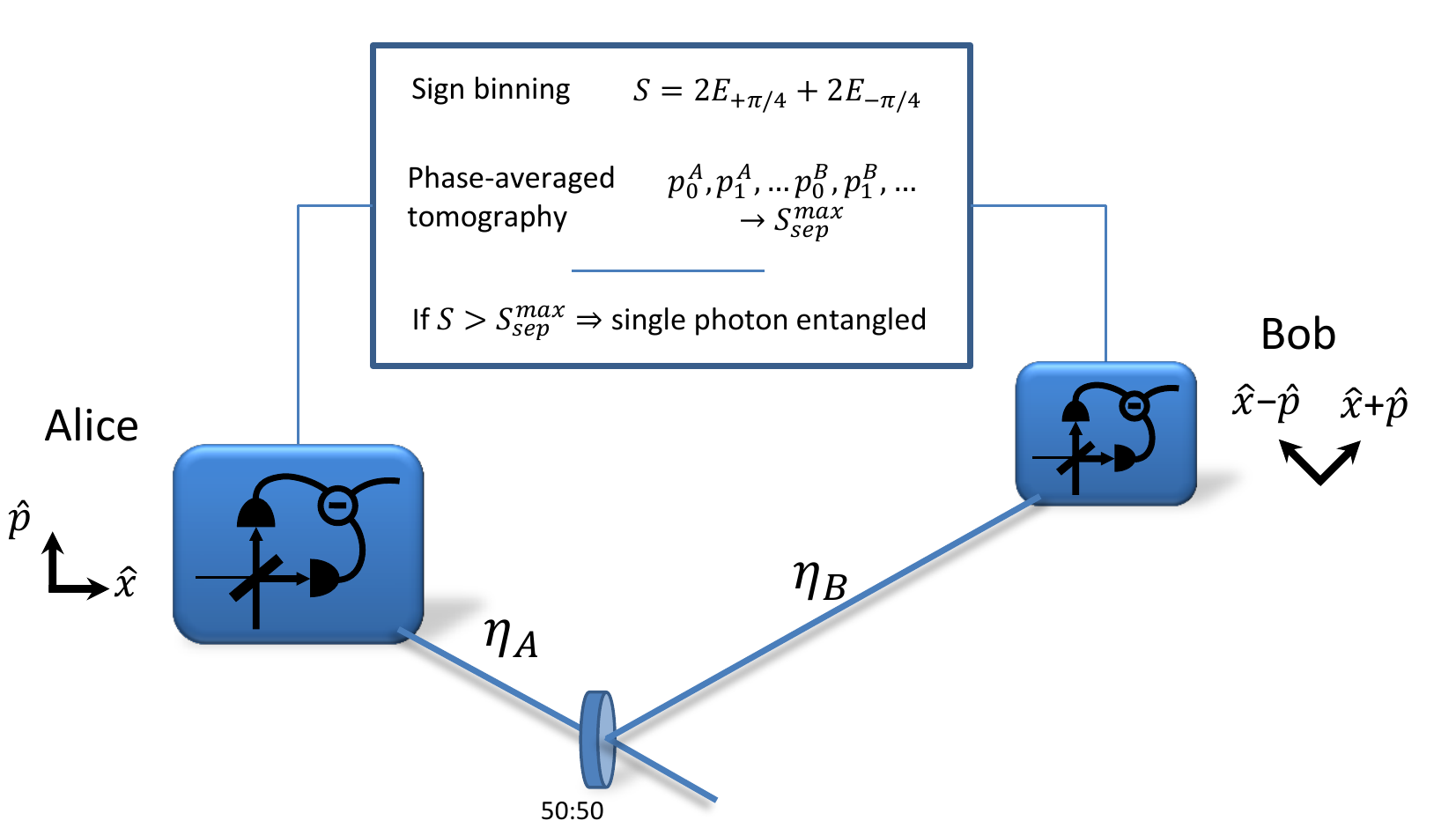}}
\caption{Principle of the entanglement witness. Single-photon entanglement is generated by impinging a single-photon state onto a 50/50 beam-splitter. Symmetric or asymmetric losses are then induced by the communication channels, with transmission efficiencies denoted $\eta_A$ and $\eta_B$. To witness the entanglement, the two distant parties, Alice and Bob, randomly choose a measurement along two quadratures, for instance $\{X,P\}$ for Alice and $\{X+P,X-P\}$ for Bob. The phase of the local oscillators are phase averaged, only the relative phase between the two detections is fixed. Sign-binning of the quadrature measurements enables then to calculate the witness parameter $S$, which has to be compared to the separability bound $S_{sep}^{max}$. This bound depends on the multi-photon components and can be optimized by using the local probabilities, which can be directly accessed from the same data thanks to the phase-averaging.}
\label{fig:general_scheme}
\end{figure}

The general principle of the witness is shown in figure \ref{fig:general_scheme}. The two distant entangled modes are detected by Alice and Bob via homodyne detection, which allows one to measure any quadrature component of the optical field, i.e. $X\,\cos(\phi)+P\,\sin(\phi)$, by varying the relative phase $\phi$ between the optical mode and the local oscillator \cite{Leonhardt}. Two phase settings are required on each side: Alice performs a measurement among two quadratures $\{X,P\}$ while Bob makes a measurement in a basis rotated by $45^\circ$ to access the quadratures $\{X+P,X-P\}$. The measurement outcomes, which are real numbers, are then sign-binned to obtain binary results $\pm1$. The scenario is thus similar to the usual Bell test where two parties can perform two possible measurements of two outcomes each. 
From the four possible combinations of quadratures, the witness parameter $S$ is finally determined from the Clauser-Horne-Shimony-Holt (CHSH) polynomial \cite{Clauser}:
\begin{equation}
\label{S}
S=E_{X,X+P}+E_{X,X-P}+E_{P,X+P}-E_{P,X-P},
\end{equation}
where the correlations are defined by $E_{a,b}=p(1,1)+p(-1,-1)-p(1,-1)-p(-1,1)$ and $p(i,j)$ are the conditional probabilities to obtain the outcomes $i$ and $j$ if the quadratures $a$ and $b$ are chosen. 

Additionally, the phase of the local oscillators can be averaged while keeping the relative phases between Alice and Bob's measurements fixed. This averaging can only lead to underestimate the entanglement as it can be realized by local operations and classical communications. The $S$ parameter reduces thus to two terms, one where the relative phase differs by $\frac{\pi}{4}$, and the other by $-\frac{\pi}{4}$:
\begin{equation}
\label{S'}
S=2E_{+\pi/4}+2E_{-\pi/4}.
\end{equation}
As shown in \cite{Morin2013}, this phase-averaging is actually crucial in the protocol as it enables us to also access the local photon-number probabilities with the same homodyne measurements. These probabilities are then used to further constrain the set of density matrices that we consider in our optimisation of the separable bounds.  

\subsection{Extremal values of the witness $S$ for entangled states}\label{qubitbound}
Sign-binning of homodyne measurement in the qubit subspace $\{|0\rangle,|1\rangle\}^{\otimes 2}$ is equivalent to a noisy spin measurement \cite{Quintino2012,Sangouard}. For instance, the operator associated to a sign-binned $X$-measurement corresponds to $\sqrt{2/\pi}\,\hat{\sigma}_x$ where $\hat{\sigma}_x$ is the standard Pauli matrix. A maximally entangled state, $\left(|1\rangle|0\rangle+|0\rangle|1\rangle\right)/\sqrt{2}$, thus leads to $S_{max}=2\sqrt{2}\,.2/\pi\simeq 1.8$, the maximal value that one can obtain using the aforementioned measurements. Note that since this value is lower than 2, a violation of the well-known local bound for the CHSH polynomial is not possible in this context. While this would have been sufficient to demonstrate entanglement, it is not necessary if the separable bound is lower.

The next question that arises is then the value of the separable bound. It can be shown that the maximal value over the set of all the separable states is equal to $S_{sep}=\sqrt{2}\,.2/\pi\simeq 0.9$ \cite{Roy2005}. In the qubit space, an observed $S$ parameter above 0.9 allows one to conclude that the two modes are entangled. Importantly, this separable bound can be optimized further if additional knowledge about the state is available, as this knowledge constrains the set of compatible separable states. 
In our case, the phase-averaged homodyne measurements provides us the local photon number distributions. These local photon-number distributions  $p_0^A,p_1^A$ (vacuum and single-photon component on Alice side) and $p_0^B,p_1^B$ (Bob side) allow us to optimize the bound, as shown now.

First, thanks to the averaging of the local phases, many off-diagonal terms of the measured state do not contribute to the measurement results. Since our goal is to reveal entanglement, it is therefore sufficient to consider density matrices of the following form in the Fock basis \cite{Chou05}:
\begin{equation}
\hat{\rho} = 
\begin{pmatrix}
p_{00} & 0 & 0 & 0 \\
0 & p_{01} & \textit{\textbf{d}} & 0 \\
0 & \textit{\textbf{d}}^* & p_{10} & 0 \\
0 & 0 & 0 & p_{11} \\
\end{pmatrix}.
\end{equation}
Then, for any state within the qubit subspace, it can be shown that the $S$ parameter can be rewritten as \cite{Morin2013}
\begin{equation}
S=\frac{16}{\pi\sqrt{2}}\Re\left[\bra{01}\hat{\rho}\ket{10}\right]=\frac{16}{\pi\sqrt{2}}\Re\left[\textit{\textbf{d}}\right]. \label{Sd}
\end{equation}

When Alice and Bob measure the value of $S$, they can also extract from the quadrature measurements the local probabilities $p_0^A$ and $p_0^B$. Hence, only a reduced set of states are compatible with these probabilities. It can be translated formally as:
\begin{itemize}
\item $p_0^A=p_{00}+p_{01}$ and $p_0^B=p_{00}+p_{10}$  (relationship between joint probabilities and local probabilities)
\item $\text{tr}[\hat{\rho}]=1$ (conservation of probabilities)
\item $\hat{\rho} \geqslant 0$ (physical state, all eigenvalues are positive), i.e.  $p_{01}p_{10}\geqslant \abs{\textit{\textbf{d}}}^2$
\item $0\leqslant p_{ij}\leqslant1$ (regular probabilities)
\end{itemize}
The maximization of $\abs{\textit{\textbf{d}}}$ under all these constraints gives the upper bound $S^\text{max}$ for the witness parameter:
\begin{equation}
S^\text{max}=\frac{16}{\pi\sqrt{2}}\begin{cases}
\sqrt{p_0^Ap_0^B}& \text{ if   $p_0^A+p_0^B\leqslant1$},\\
\sqrt{(1-p_0^A)(1-p_0^B)}& \text{ if   $p_0^A+p_0^B\geqslant1$}.
\end{cases}
\label{eq:s_max}
\end{equation}

We now derive the separable bound $S_\text{sep}^\text{max}$. Separable states remain positive under partial transposition (PPT criterion) \cite{Peres, Horodecki}. This additional constraint leads to the condition $\abs{\textit{\textbf{d}}}^2 \leqslant p_{00}p_{11}$ for separable states. Hence, the maximization of $\abs{\textit{\textbf{d}}}$ provides the maximal value of $S$ but, this time, for the separable states only:
\begin{equation}
S_\text{sep}^\text{max}=\frac{16}{\pi\sqrt{2}}\sqrt{p_0^Ap_0^B(1-p_0^A)(1-p_0^B)}.
\label{eq:s_sep_max}
\end{equation}

\subsection{Witnessing single-photon entanglement after losses}\label{lossesAnalysis}
We now study the use of the proposed witness in the case where the entangled state undergoes loss, e.g. propagates through lossy communication channels. What are the acceptable losses in this case? With the help of the analytical bounds derived previously, we detail how the proposed witness is affected. 

The situation is sketched on figure \ref{fig:general_scheme}. We consider the entanglement initially generated from an ideal single-photon state and the channel transmissions are denoted $\eta_A$ from the source to Alice and $\eta_B$ from the source to Bob. One can write the full transmission between Alice and Bob as $\eta_{AB}=\eta_A\eta_B$. After propagation, the resulting state shared by Alice and Bob can be written as:
\begin{equation}
\hat{\rho}_{AB} = \frac{1}{2}
\begin{pmatrix}
2-\eta_A-\eta_B & 0 & 0 & 0 \\
0 & \eta_A & \sqrt{\eta_A\eta_B} & 0 \\
0 & \sqrt{\eta_A\eta_B} & \eta_B & 0 \\
0 & 0 & 0 & 0 \\
\end{pmatrix}.
\label{eq:state_losses}
\end{equation}
As given by eq. (\ref{Sd}), the CHSH polynomial value $S$ can be written as:
\begin{equation}
S(\hat{\rho}_{AB})=\frac{16}{\pi\sqrt{2}}\frac{\sqrt{\eta_A\eta_B}}{2}\ .
\label{eq:S_asym}
\end{equation}
Furthermore, the local probabilities are given by:
\begin{equation}
p_1^A=\eta_A/2 \quad \textrm{and} \quad p_1^B=\eta_B/2.
\end{equation}
The maximal value of eq. (\ref{eq:s_max}) is saturated by the state given in eq. (\ref{eq:state_losses}) and the corresponding separable bound is
\begin{equation}
S_\text{sep}^\text{max}=\frac{8}{\pi\sqrt{2}}\sqrt{\eta_A \eta_B (1-\eta_A/2)(1-\eta_B/2)}\ .
\end{equation}

With this simple model in hand, one can distinguish two different experimental scenarios. First, when the source is placed on Alice's site, the losses are thus asymmetric and $\eta_A=1$ and $\eta_B=\eta_{AB}$. For this configuration, the separable bound is
\begin{equation}
S_\text{sep}^\text{max}(asym.)=\frac{4}{\pi}\sqrt{\eta_{AB}(1-\eta_{AB}/2)}\ .
\end{equation}
The second scenario places the source at an equal distance from Alice and Bob so that the state will propagate along the same distance on both arms. The two modes are thus affected by the same losses $\eta_A=\eta_B=\sqrt{\eta_{AB}}$, leading to the following separable bound:
\begin{equation}
S_\text{sep}^\text{max}(sym.)=\frac{8}{\pi\sqrt{2}}\sqrt{\eta_{AB}}(1-\sqrt{\eta_{AB}}/2)\ .
\end{equation}
In order to compare both cases, we fix the full transmission $\eta_{AB}=\eta_A\eta_B$. In other words the position of the source is changed but not the total distance between Alice and Bob. Furthermore, we note that the symmetric situation can equivalently correspond to losses on the source itself. Indeed, it is formally equivalent to attribute these losses to the two transmission channels.

\begin{figure*}[htpb!]
\centerline{
\includegraphics[width=0.85\textwidth]{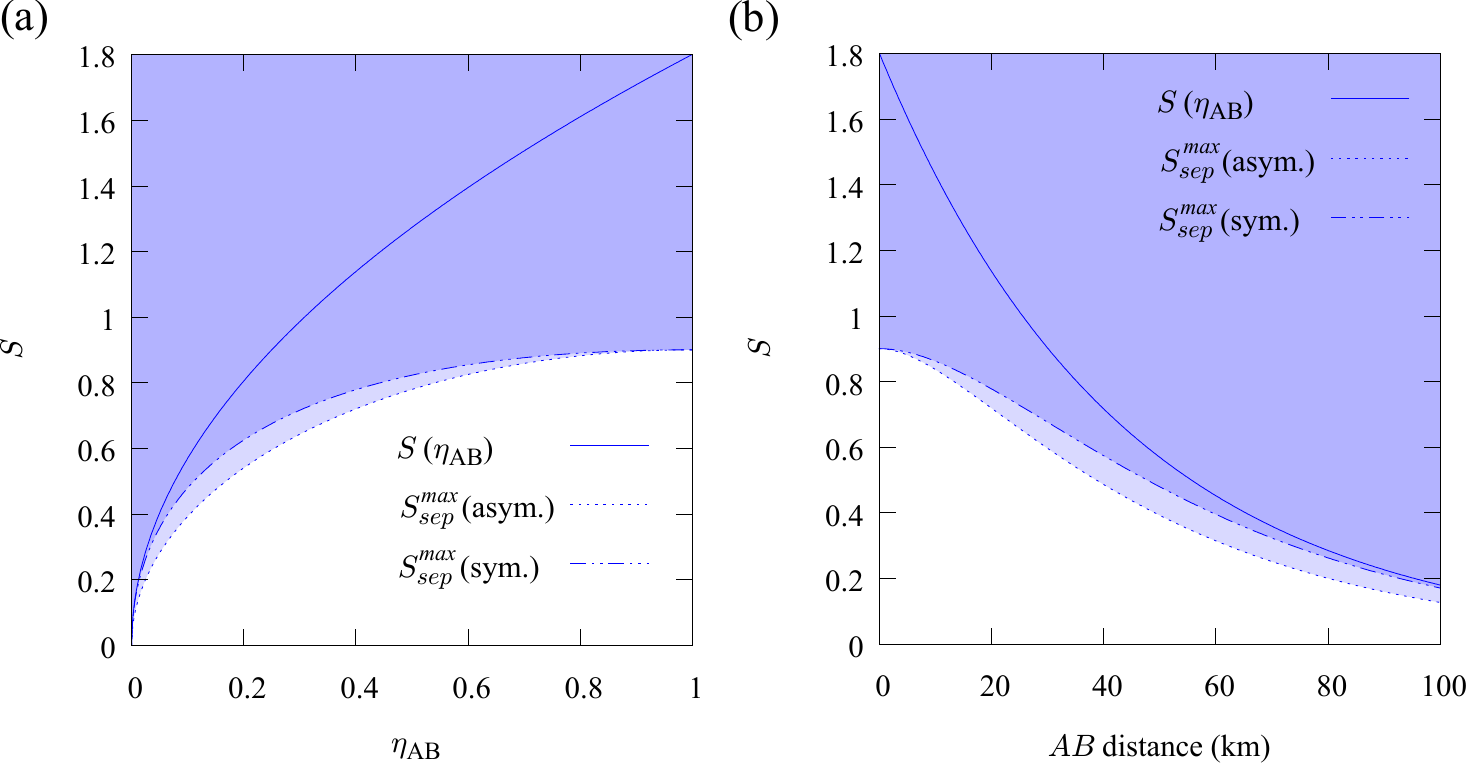}}
\caption{(a) Values of the  CHSH parameter $S$ and of the separable bounds $S_\text{sep}^{max}$ when applied to single-photon entanglement propagated through lossy communication channels. Two cases are considered: when the total losses are only on one transmission channel (asymmetric case, $\eta_A=1$ and $\eta_B=\eta_{AB}$) and when the losses are symmetric on the two channels ($\eta_A=\eta_B=\sqrt{\eta_{AB}}$). (b) The figure corresponds to the same results but with a scale given in kilometers of propagation in a fiber at telecom wavelength (0.2 dB-per-km loss).}
\label{fig:plots_model_a_as}
\end{figure*}

Figure \ref{fig:plots_model_a_as} provides the CHSH polynomial as a function of the transmission, together with the two separable bounds. As shown before, the parameter $S$ depends only on the total loss while the separable bound depends also on whether the losses are symmetric or asymmetric. As can be seen, the distance between the witness and the bound is decreasing with the losses but reaches zero only for infinite ones, meaning that in principle the witness can detect entanglement for any losses. Furthermore, we note that the distance of $S$ from the separable bound is always larger for the asymmetric case than for the symmetric one. The witness is thus slightly more efficient in this latter case.

\section{Including higher photon numbers: general case}\label{bounding}
In the previous section, the separable bound has been determined by considering that the state contains at most one photon per mode. However, states produced in a setup generally include multiphoton components that have critically to be taken into account. We present here separable bounds which can be used in this case. Namely, we provide three such expressions. Each one has different strengths, which we summarize in the last part of this section. We first present the approach we use to bound the possible effect of multiphoton components on the witness $S$.

\subsection{Bounding the Hilbert space}
When the number of photons per mode is not restricted to one, we use the local photon number distributions obtained via phase-averaged quantum state tomography to determine an upper bound on the joint probability $p_{joint}=p(n_A\geq 2 \cup n_B\geq 2)$ that at least one of the modes is populated with more than one photon. Indeed, this probability can be bounded by the local probabilities of the zero and one photon components on each side as:
\begin{equation}
\label{bound}
p_{joint}\leq p^\star,
\end{equation}
where $p^{\star}=p^A_{\geq2}+p^B_{\geq2}$, and $p^{A(B)}_{\geq2} = 1-p^{A(B)}_{0}-p^{A(B)}_{1}$, denoting the probability that one party observes at least 2 photons.

In the following, we thus present some separable bounds in terms of $p_{joint}.$ These can be re-expressed in terms of local photon distributions by substituting $p^{\star}$ for $p_{joint}$, hence slightly overestimating the bound.

\subsection{A first separable bound as a function of the local probabilities}
Following a similar argument as presented in section~\ref{qubitbound}, we provide here a separable bound for $S$ valid outside of the qubit space.

In this larger Hilbert space, $S$ can be bounded as follows (c.f. \cite{Morin2013}):
\begin{equation}
S \leq  \frac{16}{\pi\sqrt{2}}d + \frac{8}{\pi} e + \frac{8}{\pi} f + 2\sqrt{ 2} \  p_{joint},
\label{eq:S2photons}
\end{equation}
where $d =  \Re\left[\langle 01 | \hat{\rho} | 10\rangle\right]$,  $e = \Re\left[\langle 20 | \hat{\rho} | 11\rangle\right] $, and  $f = \Re\left[\langle 02 | \hat{\rho} | 11\rangle\right]$ denote different contributions to the witness. Due to the positivity of $\hat{\rho}$ and $\hat{\rho}^{T_B(0,1)}$, each of these contributions can be bounded as a function of a single density matrix variable $p_{00}$:
\ba
d^2 \leq p_{01}p_{10} \leq (p^A_0-p_{00})(p^B_0-p_{00})   \nonumber \\
d^2 \leq  p_{00}p_{11} \leq p_{00} \left[ p_{00} +1 -p^A_0 -p^B_0 + p^A_{\geq 2} + p^B_{\geq 2}\right] \nonumber \\
e^2 \leq p_{02} p_{11} \leq p^B_{\geq 2} \left[ p_{00} +1 -p^A_0 -p^B_0 + p^A_{\geq 2} + p^B_{\geq 2}\right] \nonumber \\
f^2 \leq  p_{20} p_{11} \leq p^A_{\geq 2} \left[ p_{00} +1 -p^A_0 -p^B_0 + p^A_{\geq 2} + p^B_{\geq 2}\right]  
\ea
The maximum value of $S_{sep}$ can thus be obtained by optimizing Eq. ~\eqref{eq:S2photons} over the $p_{00}$ variable. Recall that here, we do not impose the state $\hat{\rho}$ to be fully PPT, but only PPT within the 0/1 subspace. This allows us to verify the presence of entanglement in the single-photon subspace~\cite{Morin2013}.
 
For small $p^{A(B)}_{\geq 2}$, the choice $p_{00}^c = p^A_0 p^B_0/z$ is optimal, where $z = 1+p^A_{\geq 2} + p^B_{\geq 2}$. This gives the following separable bound:
\begin{equation}
\begin{split}
S_{sep}^{max} =& \frac{16}{\pi\sqrt{2} }\sqrt{p^A_0 p^B_0 \left(1-\frac{p^B_0}{z} \right) \left(1-\frac{p^A_0}{z} \right) } \\
&+ \frac{8}{\pi} \left( \sqrt{p^A_{\geq2}} + \sqrt{p^B_{\geq 2}} \right)\sqrt{z + \frac{p^A_0 p^B_0}{z} - p^A_0 - p^B_0} \\
&+ 2\sqrt{ 2} \  p_{joint}.
\end{split}\label{eq:shortcutBound}
\end{equation}

One can verify that this expression reduces to the qubit bound given by Eq. ~\eqref{eq:s_sep_max} in the case $p^{A(B)}_{\geq 2} = 0$. Equation ~\eqref{eq:shortcutBound} provides an analytical estimation of the value of the witness needed to demonstrate single-photon entanglement as a function of the local observed probabilities. We emphasize that this bound is also valid in presence of multiphoton components.

\subsection{Separable bound as a function of $p_{joint}$}\label{pstarBound}
Here we derive a separable bound which only depends on the $p_{joint}$ variable. Writing matrices $M$ and $N$ such that 
\begin{eqnarray}
\text{Tr}(M\hat{\rho}) &=& \frac{16}{\pi\sqrt{2}}  \Re\Big[\langle 01 | \hat{\rho} | 10\rangle\Big] +\nonumber \\
&& \frac{8}{\pi} \left(\Re\Big[\langle 20 | \hat{\rho} | 11\rangle\Big]+\Re\Big[\langle 02 | \hat{\rho} | 11\rangle\Big]  \right) \\
\text{Tr}(N\hat{\rho}) &=&  \langle 00 | \hat{\rho} | 00\rangle+\langle 01 | \hat{\rho} | 01\rangle \nonumber \\
&&+\langle 10 | \hat{\rho} | 10\rangle+\langle 11 | \hat{\rho} | 11\rangle,
\end{eqnarray} 
the maximum separable value of Eq.~\eqref{eq:S2photons} given $p_{joint}$ can be found by maximizing $\text{tr}(M\hat{\rho})+2\sqrt{2}p_{joint}$ under the constraint that $\hat{\rho} \geq 0$, $\text{tr}(\hat{\rho})    \leq 1$, $\hat{\rho}^{T_B (0,1)} \geq 0$ and $\text{tr}(N\hat{\rho}) = 1- p_{joint}$.

Any matrices $A$ and $B$, and variables $\lambda$ and $\mu$ that satisty  $ A + B^{T_B (0,1)} -\mu N - \lambda I = - M$, $A\geq 0$, $B\geq 0$ provide an upper bound on the result of this optimization. Indeed, these constraints guarantee that
\begin{eqnarray} \label{SDP}
\text{tr}(M\hat{\rho}) &=& \text{tr}[( \lambda I+\mu N  -A - B^{T_B (0,1)} )\hat{\rho}]  \nonumber \\
&=& \lambda \text{tr}(\hat{\rho})   + \mu \text{tr} (N\hat{\rho}) - \text{tr}(A\hat{\rho}) - \text{tr}(B\hat{\rho} ^{T_B (0,1)} ) \nonumber \\
&\leq& \lambda + \mu(1-p_{joint}).
\end{eqnarray}  
We describe, in the appendix, matrices $A$ and $B$ that satisfy these constraints for $p_{joint} \leq 1/2$, $\lambda=\frac{2}{\pi}\sqrt{\frac{2}{p_{joint}}}x_+$, $\mu=(\frac{2}{\pi}\sqrt{2}x_+^2-\lambda)/(1-p_{joint})$ and $x_\pm=\sqrt{1-p_{joint}} \pm \sqrt{p_{joint}}$. This gives the following maximum for the separable bound:
\begin{equation}\label{eq:pstarBound}
S_{sep}^{max} = 2\sqrt{2} \left[\frac{1}{\pi}  (\sqrt{1-p_{joint}} + \sqrt{p_{joint}})^2  + p_{joint} \right]
 \end{equation}
One can check that this bound is achievable for all $p_{joint}\leq1/2$ by some quantum states $\hat{\rho}$ which are PPT in the single-photon subspace. This guarantees that the bound is tight as a function of $p_{joint}$. However, this bound does not take into account the local probabilities.

\subsection{A refined semidefinite bound taking advantage of local probabilities}\label{refinement}
In Ref. ~\cite{Morin2013}, a semidefinite program (SDP) is presented to compute separable bounds on $S$ as a function of the local photon number probabilities. Here, we provide a refined version of this program including two improvements.

The first improvement is to express $p_{joint}$ in~\eqref{eq:S2photons} as a function of the density matrix elements rather than bounding it according to Eq. ~\eqref{bound}. This allows us to perform the optimization of $S$ across all terms together.

The second step is to take into account all information about the local probability distributions. This can be achieved by using the Frechet Inequalities~\cite{Frechet35}. In the form of the disjunction\footnote{The conjunction and disjunction form of the Frechet inequalities can be found to be equivalent, so we use only one form.}, these inequalities can be expressed as:
\begin{eqnarray}
\hspace{-0.5cm}\max(0, p(A) +p(B) -1)&\leq&  P(A \cap B)\nonumber\\ &&\leq \min(p(A), p(B)).
\end{eqnarray}
Here, $A(B)$ refers to any set that includes at least one photon number on Alice's (Bob's) side. For instance, in the case that probabilities up to 1 photon component are observed, the possible choices for $A$ and $B$ consist of any nonempty combination from $  \{\text{0 photon, 1 photon, more than 1 photon}\}$. This gives us a set of $(2^3-1)$ by $(2^3-1)$ separate Frechet inequalities.

Adding the usual conditions to the two we just mentioned leads to the following formulation for the refined bound:
\begin{eqnarray}
\label{optimisation_enhanced}
 \max\ \ &&  S(p^A_0,p^A_1  ,p^B_0 ,p^B_1)    \\  
 \text{\ \ s.t.\ \ } && \hat{\rho} \geq 0 \nonumber\\
 &&{\rm tr} (\hat{\rho}) \leq 1\nonumber\\
 && \hat{\rho}^{T_b(0,1)} \geq 0 \nonumber\\
&&  P(A \cap B) \geq \max [\  0, p(A) +p(B) -1 \ ],  \   \forall A,B \nonumber\\
 &&   P(A \cap B) \leq \min [\  p(A), p(B) \ ],    \   \forall A,B \nonumber
 \end{eqnarray}
The program described in~\cite{Morin2013} can be seen as a relaxation of this one.

As presented here, it should be clear that the program~\eqref{optimisation_enhanced} can be extended to take into account additional local photon numbers. In this case, the expression~\eqref{eq:S2photons} needs to be modified to fit the new considered Hilbert space. Similarly, the definition of $p_{joint}$ can be adapted. However, the program remains the same. This presents the possibility of enhancing the bounds by taking into account additional information. We come back to this possibility in the experimental part of this paper.

Finally, we note that uncertainties in the local probabilities can be taken into account in this method by following the same procedure than presented in~\cite{Morin2013}.

\subsection{Comparison of the separable bounds}
Until now, we have presented four separable bounds for the witness. Let us briefly highlight their differences and mention the context in which one could be interested in using each of them.

The first bound, given in Eq.~\eqref{eq:s_sep_max}, is valid only for qubit states, and is thus not applicable in practice. However, it takes advantage of the observed local photon number distributions. This is the bound we used in section~\ref{lossesAnalysis} to first illustrate the effect of losses on the witness.

The second bound, given in Eq.~\eqref{eq:shortcutBound}, also takes advantage of the knowledge of the photon number distributions and applies outside of the qubit space. However, one can check that this bound is not always tight. This comes from the fact that only some of the Frechet inequalities were taken into account in its derivation. Moreover, this bound can be very sensitive to uncertainties in the local probabilities, making it hardly applicable in practice. Nevertheless, it can be useful to estimate quickly the value of the bound that can be derived from Eq.~\eqref{optimisation_enhanced}.

The third bound, given in Eq.~\eqref{eq:pstarBound} is tight as a function of $p_{joint}$ alone. It behaves well in presence of uncertainties, but does not take advantage of the knowledge of the local photon number probabilities.

The fourth bound, given in Eq.~\eqref{optimisation_enhanced} is expressed as a semidefinite program. It does not assume a qubit structure and computes the tightest separable bound compared to all other methods by taking all physical constraints into account. Moreover, since it includes an exact modelization of the underlying quantum state, it behaves well in presence of uncertainties on the local probabilities. This is thus the kind of bound that we use in the next section to analyse the experimental data.

\section{Experimental implementation of the witness}\label{manip}
In this section, we present the experimental setup, including the single-photon source used for the entanglement generation and the practical details for implementing the witness.  The experimental behavior of the witness with losses is finally given. 

\subsection{Single-photon source and entanglement}
The single-photon source is based on a type-II optical parametric oscillator pumped far below threshold by a continuous-wave frequency-doubled Nd:YAG laser at 532 nm \cite{Laurat04}. The frequency-degenerate signal and idler modes are orthogonally polarized and can be easily separated. The detection of a single-photon on one mode then heralds the preparation of a single photon in the other one \cite{Hong,D'Auria11,D'Auria12}. Importantly, the photon is generated in a very well-defined spatiotemporal mode due to the OPO cavity. Experimental details, including the filterings required in the conditioning path and the definition of the temporal mode, have been presented elsewhere \cite{Morin2012, Morin2013b,JOVE}. In the current experiment, the heralding efficiency, i.e. the single-photon component at the output of the OPO, is equal to 90 \% and the two-photon component is limited to a few percents. If one includes the total propagation and detection losses, the single-photon component reaches 68$\pm$2\%. The initial effective transmission $\eta_{AB}$ is thus $\sim 0.68.$ 

Entanglement is obtained by impinging the heralded single-photon state on a balanced polarizing beam splitter. To check the entanglement, the two modes are then directed to two homodyne detections, as shown in figure \ref{fig:general_scheme}. By using the previous notations, without introducing additional communication channel losses, $p_1^{A}=p_1^{B}=0.34\pm0.01$, in comparison to $p_1^{A}=p_1^{B}=0.5$ for entanglement generated from an ideal single-photon source. In the following, we include additional losses to decrease the transmission, in a symmetric or asymmetric way.

\begin{figure*}[htpb!]
\centerline{
\includegraphics[width=0.9\textwidth]{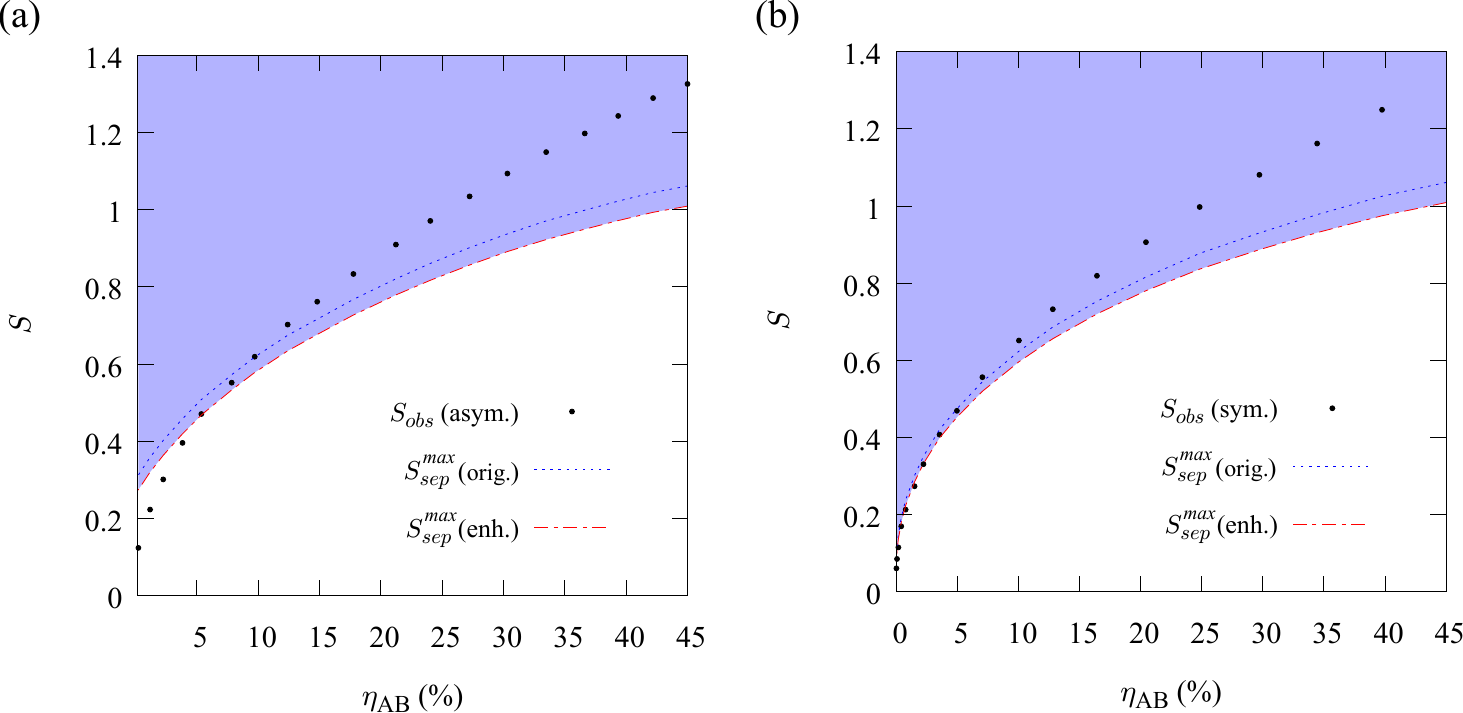}}
\caption{Experimental results. In the asymmetric case (a), additional losses are applied on one channel whereas in the symmetric case (b), losses are applied equally on both ones. The results are given as a function of the overall transmission $\eta_{AB}=\eta_A\eta_B$. The figures provide the measured CHSH values $S_{obs}$ (the size of the points accounts for statistical errors) together with two separable bounds determined following the program given in section \ref{refinement}. $S_{sep}^{max}(orig.)$ takes into account the multi-photon components up to 1 photon as presented in~\cite{Morin2013}. $S_{sep}^{max}(enh.)$ corresponds to an enhanced separable bound, which takes into account in the optimization the two-photon components as additional constraints.}
\label{fig:results}
\end{figure*}

\subsection{Witnessing entanglement: practical realization}
To perform the homodyne detections, a bright beam impinges on the balanced polarizing beam splitter mentioned above in order to distribute the two required local oscillators. The classical and quantum channel thus have orthogonal polarizations but the same spatial modes up to the detections. This configuration allows one to easily adjust the relative phase between the two detections by choosing an appropriate elliptical polarization for the bright beam before the splitting \cite{Laurat05,Morin2013}. By sweeping also its phase, both homodyne detections have thus a fixed relative phase but are locally phase-averaged, as required.

We now detail the full experimental procedure for implementing the proposed witness. The steps are as follows:
\begin{itemize}
\item{\textit{Acquiring homodyne data.} Phase-averaged homodyne tomography is performed on both modes. Four relative phase settings are required but phase-averaging enables to reduce them to two, i.e. $\pm\pi/4$. The recorded data are then used for the next steps.}
\item{\textit{Extracting the local probabilities.} The local photon-number distributions are extracted from the previous data. Importantly, no additional measurements are required. The estimation is obtained thanks to pattern functions that relax any assumptions on the size of the Fock space \cite{Munroe95}.}
\item{\textit{Determining the separable bound.} The local probabilities are used to constrain the set of separable states and calculate the separable bound following the program given in eq. \ref{optimisation_enhanced}.  }
\item{\textit{Calculating the $S$ parameter.} The homodyne data are sign-binned and the $S$ parameter is then determined from Eq.~\eqref{S}. If $S$ is above the separable bound, the bipartite state is entangled.}
\end{itemize}

\subsection{Tunable losses}
We now turn to the study of the effect of losses on the proposed witness. Losses have been simulated here by changing the temporal modes. Indeed, the experiment is based on continuous-wave homodyne detection, i.e. the quadrature measurement is a continuous signal $x(t)$. In order to measure the mode in which our state lies, a temporal filtering is required, leading to $x_\psi=\int\psi(t)x(t)dt$. The optimal temporal mode $\psi(t)$ contains the generated state and all the other orthogonal modes contain a vacuum state \cite{Morin2013b}. We can thus generate controlled and tunable losses by mismatching the temporal mode we chose and the optimal one. The overlap $\int\psi(t+\tau)\psi(t)dt$ provides the additional losses $\eta(\tau)$ on each channel. As done on the same raw data, the original state is always the same, only the losses are tuned by this procedure. 

\subsection{Results}
Experimental results are displayed on Figure \ref{fig:results}. The measured CHSH parameter is given for  different values of losses, together with the corresponding separable bounds. In the asymmetric case, the losses have been increased on one of the homodyne detections, while for the symmetric case, the losses are generated equally on both of them. However, as we did in the model, we can compare both situations in a relevant fashion only if we consider the full losses: we estimate the local losses with the help of the experimental vacuum components $p_0^A=1-\eta_A$ and $p_0^B=1-\eta_B$, and then obtain the corresponding overall transmission $\eta_{AB}=\eta_A\eta_B$.

The obtained results are in very good agreement with the expected behavior, i.e. $S\propto \sqrt{\eta_{AB}}$, and show that the bound of the single-photon entanglement witness can be violated unless very significant losses are incurred. The figure provides two separable bounds determined following the program given in \ref{refinement}. The first one takes as constraint the local probabilities up to 1 photon, as considered in \cite{Morin2013}, while the second one considers the two-photon component. Clearly, more losses can be tolerated thanks to this enhanced separable bound. In the asymmetric case, the limit is pushed experimentally from 90\% to 95\% (corresponding to around 65 km of fiber at telecom wavelength if one starts with ideal single-photon) while in the symmetric case a rise from 95\% to 97\% (77 km of fiber) is obtained. This small difference between the symmetric and asymmetric cases can be explained by the higher photon number component and the sensitivity of the bound to this parameter. Indeed, in the asymmetric case, the mode that does not experience loss keeps a larger two-photon component, which allows for a separable state with higher $S$ parameter.

\section{Conclusion}\label{Discussion}

In summary, we have presented a detailed analysis of the scheme to witness single-photon entanglement based only on local homodyne measurements proposed in Ref. \cite{Morin2013}. The effect of losses have been considered and our investigation has shown the robustness of this hybrid witness. Even with communication channel losses of around 95\%, entanglement can still be experimentally witnessed, whether in the symmetric or asymmetric case. The separable bound has been optimized by including local photon number distributions up to two photons. Indeed, the main contribution outside the qubit subspace comes from this component. These results confirm the efficiency of the witness and its relevance as an operational test for large-scale networks relying on single-photon entanglement.

\section*{Acknowledgments}
We would like to thank Valentina Caprara Vivoli, Valerio Scarani and Pavel Sekatski for discussions and comments on the paper. We acknowledge support by the ERC Starting Grant HybridNet, the ERA-NET CHIST-ERA under the QScale project, the Swiss NCCR QSIT, the Swiss National Science Foundation SNSF (grant PP00P2$\_$150579), the European Commission (IP SIQS), the Singapore National Research Fund and the Ministry of Education  (partly through the Academic Research Fund Tier 3 MOE2012-T3-1-009). Julien Laurat is a member of the Institut Universitaire de France.

\section*{Appendix}
Here are the matrices we use to derive the analytical bound in Section \ref{pstarBound}. The matrices are expressed in the natural basis for photon numbers, i.e. $\{\ket{00},\ket{01},\ket{02},\ket{10},\ket{11},\ket{12},\ket{20},\ket{21},\ket{22}\}$.

\begin{align}
A &= \left [ \begin{array}{ccccccccc} 
0&0&0&0&0&0&0&0&0 \\ 
0&\lambda+\mu&0&m &0&0&0&0&0\\ 
0&0&\lambda&0&-\frac{4}{\pi}&0&0&0&0 \\ 
0&m &0&\lambda+\mu&0&0&0&0&0 \\ 
0&0&-\frac{4}{\pi}&0&\ell&0&-\frac{4}{\pi}&0&0 \\ 
0&0&0&0&0&\lambda&0&0&0 \\ 
0&0&0&0&-\frac{4}{\pi}&0&\lambda&0&0 \\ 
0&0&0&0&0&0&0&\lambda&0 \\ 
0&0&0&0&0&0&0&0&\lambda \\ 
\end{array} \right]\nonumber,\\
B &= (\lambda+\mu-\ell)\left [ \begin{array}{ccccccccc} 
x_+^2/x_-^2&0&0&0&-x_+/x_-&0&0&0&0 \\ 
0&0&0&0&0&0&0&0&0 \\ 
0&0&0&0&0&0&0&0&0 \\ 
0&0&0&0&0&0&0&0&0 \\ 
-x_+/x_-&0&0&0&1&0&0&0&0 \\ 
0&0&0&0&0&0&0&0&0 \\ 
0&0&0&0&0&0&0&0&0 \\ 
0&0&0&0&0&0&0&0&0 \\ 
0&0&0&0&0&0&0&0&0 \\ 
\end{array} \right]\nonumber,
\end{align}
where
\begin{equation}
\ell=\frac{8\sqrt{2p_{joint}}}{\pi x_+},\quad\text{ and }\quad m=\frac{x_+}{x_-}(\lambda+\mu-\ell)-\frac{4\sqrt{2}}\pi.\nonumber
\end{equation}


\begin{thebibliography}{10}

\bibitem{Loock2011} P. van Loock, Optical hybrid approaches to quantum information, Laser and Photon. Rev. \textbf{5}, 167-200 (2011).

\bibitem{Furusawa2011} A. Furusawa and P. van Loock, Quantum teleportation and entanglement (Wiley-VCH, Weinheim, 2011).

\bibitem{Sangouard10} N. Sangouard, C. Simon, N. Gisin, J. Laurat, R. Tualle-Brouri and  Ph. Grangier, Quantum repeaters with entangled coherent states, JOSA B \textbf{27}, A137-A145 (2010).

\bibitem{Brask10} J.B. Brask, I. Rigas, E.S. Polzik, U.L. Andersen and A.S. Sorensen, Hybrid long-distance entanglement distribution protocol, Phys. Rev. Lett. \textbf{105}, 160501 (2010).

\bibitem{hybrid1} H. Jeong, A. Zavatta, M. Kang, S.W. Lee, L.S. Constanzo, S. Grandi, T.C. Ralph and M. Bellini, Generation of hybrid entanglement of light, Nat. Photonics \textbf{8}, 564-569 (2014).

\bibitem{hybrid2} O. Morin, K. Huang, J. Liu, H. Le Jeannic, C. Fabre and J. Laurat, Remote creation of hybrid entanglement between particle-like and wave-like optical qubits, Nat. Photonics \textbf{8}, 570-574 (2014).

\bibitem{DLCZ} L.M. Duan, M.D. Lukin, J.I. Cirac and P. Zoller, Long-distance quantum communication with atomic ensembles and linear optics, Nature \textbf{414}, 413-418 (2001).

\bibitem{vanEnk05} S.J. van Enk, Single-particle entanglement, Phys. Rev. A \textbf{72}, 064306 (2005).

\bibitem{Bjork01} G. Bj\"ork, P. Jonsson and L.L. S\'anchez-Soto, Single-particle nonlocality and entanglement with the vacuum, Phys. Rev. A \textbf{64}, 042106 (2001).

\bibitem{HessmoPRL2004} B. Hessmo, P. Usachev, H. Heydari and G. Bj\"ork, Experimental demonstration of single photon nonlocality, Phys. Rev. Lett \textbf{92}, 180401 (2004).

\bibitem{Morin2013} O. Morin, J-D. Bancal, M. Ho, P. Sekatski, V. D'Auria, N. Gisin, J. Laurat and N. Sangouard, Witnessing trustworthy single-photon entanglement with local homodyne measurements, Phys. Rev. Lett. \textbf{110}, 130401(2013).

\bibitem{Babichev2004} S.A. Babichev, J. Appel and A.I. Lvovsky, Homodyne tomography characterization and nonlocality of a dual-mode optical qubit, Phys. Rev. Lett. \textbf{92}, 193601 (2004).

\bibitem{Chou05} C.W. Chou, H. de Riedmatten, D. Felinto, S.V. Polyakov, S.J. van Enk and H.J. Kimble, Measurement-induced entanglement for excitation stored in remote atomic ensembles, Nature \textbf{438}, 828-832 (2005).

\bibitem{Laurat2007} J. Laurat, K.S. Choi, H. Deng, C.W. Chou and H.J. Kimble, Heralded entanglement between atomic ensembles: preparation, decoherence, and scaling, Phys. Rev. Lett. \textbf{99}, 180504 (2007).

\bibitem{LauratNJP} J. Laurat, C.W. Chou, H. Deng, K.S. Choi, D. Felinto, H. de Riedmatten and H.J. Kimble, Towards experimental entanglement connection with atomic ensembles in the single excitation regime, New J. Phys. \textbf{9}, 207 (2007).

\bibitem{Bennet12} A. J. Bennet, D.A. Evans, D.J. Saunders, C. Branciard, E.G. Cavalcanti, H.M. Wiseman and G.J.  Pryde, Arbitrarily loss-tolerant Einstein-Podolsky-Rosen steering allowing a demonstration over 1 km of optical fiber with no detection loophole, Phys. Rev. X \textbf{2} 031003 (2012).

\bibitem{Cavalcanti09} E.G. Cavalcanti, P.D. Drummond, H.A. Bachor,  M.D. Reid, Spin entanglement, decoherence and Bohm's EPR paradox, Opt. Express \textbf{17} 18693 (2009)

\bibitem{Kimble08} H.J. Kimble, The quantum internet, Nature \textbf{453}, 1023-1030 (2008).

\bibitem{Sangouard11} N. Sangouard, C. Simon, H. de Riedmatten and N. Gisin, Quantum repeaters based on atomic ensembles and linear optics, Rev. Mod. Phys. \textbf{83}, 33-80 (2011).

\bibitem{Leonhardt} U. Leonhardt, Measuring the quantum state of light (Cambridge University Press, Cambridge, 1997).

\bibitem{Clauser} J.F. Clauser, M. Horne, A. Shimony and R.A. Holt, Proposed experiment to test local hidden-variable theories, Phys. Rev. Lett \textbf{23}, 880-884 (1969).

\bibitem{Quintino2012} M.T. Quintino, M. Araujo, D. Cavalcanti, M. Franca Santos and M. Terra Cunha, Maximal CHSH violations with low efficiency photodetection and homodyne measurements, J. Phys. A: Math. Theor. \textbf{45}, 215308 (2012).

\bibitem{Sangouard} N. Sangouard, J-D. Bancal, N. Gisin, W. Rosenfeld, P. Sekatski, M. Weber and H. Weinfurter, Loophole-free Bell test with one atom and less than one photon on average, Phys. Rev. A \textbf{84}, 052122 (2011).

\bibitem{Roy2005} S.M. Roy, Multipartite separability inequalities exponentially stronger than local reality inequalities, Phys. Rev. Lett \textbf{94}, 010402 (2005).

\bibitem{Peres} A. Peres, Separability criterion for density matrices, Phys. Rev. Lett. \textbf{77}, 1413 (1996).

\bibitem{Horodecki} M. Horodecki, P. Horodecki and R. Horodecki, Separability of mixed states: necessary and sufficient conditions, Phys. Lett. A \textbf{223}, 1 (1996).

\bibitem{Laurat04} J. Laurat, T. Coudreau, G. Keller, N. Treps and C. Fabre, Compact source of Einstein-Podolsky-Rosen entanglement and squeezing at very low noise frequencies, Phys. Rev. A \textbf{70}, 042315 (2004).

\bibitem{Hong} C.K. Hong and L. Mandel, Experimental realization of a localized one-photon state, Phys. Rev. Lett. \textbf{56}, 58-60 (1986).

\bibitem{D'Auria11} V. D'Auria, N. Lee, T. Amri, C. Fabre and J. Laurat, Quantum decoherence of single-photon counters, Phys. Rev. Lett. \textbf{107}, 050504 (2011).

\bibitem{D'Auria12} V. D'Auria, O. Morin, C. Fabre and J. Laurat, Effect of the heralding detector properties on the conditional generation of single-photon states, Eur. Phys. Journ. D \textbf{66}, 249 (2012).

\bibitem{Morin2012} O. Morin, V. D'Auria, C. Fabre and J. Laurat, High-fidelity single-photon source based on a type-II optical parametric oscillator, Optics Lett.\textbf{37}, 3738-3740 (2012).

\bibitem{Morin2013b} O. Morin, C. Fabre and J. Laurat, Experimentally accessing the optimal temporal mode of traveling quantum light states, Phys. Rev. Lett. \textbf{111},213602 (2013).

\bibitem{JOVE} O. Morin, J. Liu, K. Huang, F. Barbosa, C. Fabre and J. Laurat, Quantum state engineering of light with continuous-wave optical parametric oscillators, J. Vis. Exp. \textbf{87}, e51224 (2014). 

\bibitem{Laurat05} J. Laurat, T. Coudreau, G. Keller, N. Treps and C. Fabre, Effects of mode coupling on the generation of quadrature Einstein-Podolsky-Rosen entanglement in a type-II optical parametric oscillator below threshold, Phys. Rev. A \textbf{71}, 022313 (2005).

\bibitem{Munroe95} M. Munroe, D. Boggavarapu, M.E. Anderson and M.G. Raymer, Photon-number statistics from the phase-averaged quadrature-field distribution: Theory and ultrafast measurement, Phys. Rev. A \textbf{52}, R924(R) (1995).

\bibitem{Frechet35} M. Fr\'{e}chet, G\'{e}n\'{e}ralisations du th\'{e}or\`{e}me des probabilit\'{e}s totales. \textit{Fundamenta Mathematica} \textbf{25}, 379-387 (1935).





\end{thebibliography}
\end{document}